\documentclass[conference]{IEEEtran}

\usepackage{url}

\usepackage[noend]{algpseudocode}
\usepackage{algorithm}
\usepackage{algpseudocode}

\usepackage{colortbl}
\usepackage{xcolor}

\usepackage{graphicx}
\usepackage{multirow}
\usepackage{subcaption}

\usepackage{cite}
\usepackage{amsmath,amssymb,amsfonts}
\usepackage{textcomp}
\usepackage{enumitem}

\usepackage[utf8]{inputenc}
\usepackage[T1]{fontenc}
\usepackage{listings}

\lstdefinestyle{mystyle}{
  language=C++,
  frame=tb,
  numbers=left,
  numberstyle=\color{gray},
  keywordstyle=\color{blue},
  commentstyle=\color{green},
  stringstyle=\color{red},
  basicstyle=\ttfamily\footnotesize,
  breaklines=true,
  breakatwhitespace=false,
  showspaces=false,
  showstringspaces=false,
  tabsize=4,
  captionpos=b,
  escapeinside={\%*}{*)},      %
  morekeywords={pragma, omp, parallel}  %
}

\usepackage{stfloats}

\usepackage{hyperref}
\usepackage{cleveref} %
\usepackage{float}
\usepackage{booktabs}
\usepackage{flushend}
\hypersetup{hidelinks,hypertexnames=false}
\makeatletter
\providecommand{\theHALG@line}{\thealgorithm.\arabic{ALG@line}}
\renewcommand{\theHALG@line}{\thealgorithm.\arabic{ALG@line}}
\makeatother

\begin{document}

\title{HeteroTask: Multi-Variant Task Scheduling and Execution on Distributed Heterogeneous Clusters}

\author{%
\IEEEauthorblockN{%
Hongzheng Tian\IEEEauthorrefmark{1}\IEEEauthorrefmark{2},
Alok Mishra\IEEEauthorrefmark{2},
Haocheng Xu\IEEEauthorrefmark{1},
Dejan Milojicic\IEEEauthorrefmark{2}, and
Sitao Huang\IEEEauthorrefmark{1}}
\IEEEauthorblockA{\IEEEauthorrefmark{1}University of California, Irvine, Irvine, CA, USA}
\IEEEauthorblockA{\IEEEauthorrefmark{2}Hewlett Packard Enterprise, Milpitas, CA, USA}
\IEEEauthorblockA{\{hongzhet, haochx5, sitaoh\}@uci.edu; \{alok.mishra, dejan.milojicic\}@hpe.com}}

\maketitle

\begin{abstract}
Heterogeneous systems that combine CPUs and GPUs across multiple servers offer substantial compute capability but complicate scheduling. For DAG-based applications with multiple task implementations, effective execution requires coordinating variant selection, device placement, and data movement. Offline planning can exploit global dependency information, yet plans derived from isolated kernel profiles may diverge from execution because of runtime overheads.

We introduce HeteroTask, a framework integrating kernel-wise profiling, PEFT-based multi-variant planning, host-buffer lifetime modeling, and schedule-driven MPI execution.
A Bayesian-optimization loop calibrates empirical overhead corrections for repeated, fixed-DAG workflows on exclusively allocated nodes.
We evaluate five benchmark and workflow-derived applications on up to six 8-GPU nodes.
On two 8-GPU nodes, HeteroTask reduces application-level completion time by 19.7--72.1\% relative to the compared StarPU implementations.
For Canny Edge Detection, calibration reduces task-window makespan by 25.0\% relative to planning without overhead corrections.
Combined with CPU/GPU variant selection, calibration reduces it by 58.9\% relative to an uncalibrated GPU-only execution.
\end{abstract}

\section{Introduction}\label{sec:Introduction}
Heterogeneous systems combine CPUs and GPUs to accelerate applications with different computational requirements~\cite{mittal2015survey,alaei2025survey}. For a workflow represented as a directed acyclic graph (DAG), each task may have several implementations with different costs. Scheduling must decide which implementation to run, where, and how to move its inputs across devices and servers. These decisions interact: a fast kernel can be a poor placement if its transfers delay downstream tasks or its intermediate buffers increase memory pressure.

\begin{figure}[ht]
    \centering
    \begin{subfigure}[t]{0.9\linewidth}
        \centering
        \includegraphics[width=0.9\linewidth]{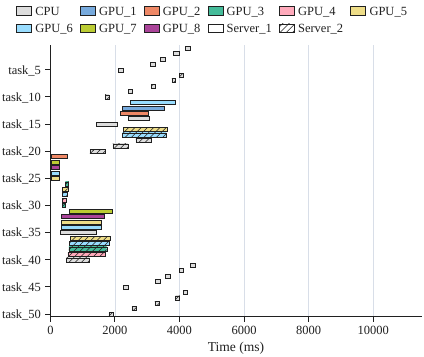}

        \caption{PEFT schedule without overhead calibration.}
        \label{fig:intro_gantt_gap_a}
    \end{subfigure}

    \vspace{11pt}

    \begin{subfigure}[t]{0.9\linewidth}
        \centering
        \includegraphics[width=0.9\linewidth]{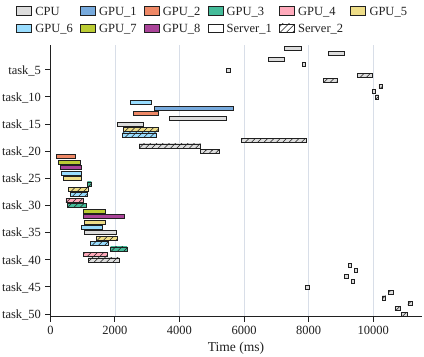}

        \caption{Actual runtime execution by applying the same schedule.}
        \label{fig:intro_gantt_gap_b}
    \end{subfigure}
    \caption{Uncalibrated plan/execution gap on two 8-GPU nodes. Total spans are 4.584 s (plan) and 11.334 s (execution), including two virtual tasks omitted from the 50 visible rows. Rows are ordered by task name, not topological ID.}
    \label{fig:intro_gantt_gap}
    \vspace{-10pt}
\end{figure}

When the full DAG and input shapes are known, offline planning can coordinate these choices across the application.
Its effectiveness, however, depends on how well the planned execution survives contact with the runtime.
Isolated kernel measurements do not capture every launch, synchronization, and data-management cost.
Fig.~\ref{fig:intro_gantt_gap} illustrates this gap for ten batched Canny Edge Detection (CED) instances on two 8-GPU nodes.
A plan based on isolated profiles and a zero-overhead vector predicts 4.584~s, whereas its execution takes 11.334~s.
Closing this gap requires attention to both the planning model and the executor that realizes its decisions.

Offline heuristics such as HEFT and PEFT exploit complete dependency and cost information~\cite{heft,peft}. Task runtimes such as StarPU, Taskflow, PaRSEC, and Legion provide mechanisms for heterogeneous execution and runtime placement~\cite{starpu,taskflow,parsec,legion}. Static knowledge and online adaptation offer complementary advantages~\cite{agullo2016static,chamorro2025online}. HeteroTask focuses on repeated workflows whose DAG structure, task shapes, and available implementations are fixed, on exclusively allocated nodes. Under these conditions, an offline plan can be reused, and a schedule-driven runtime can preserve its placement decisions while resolving dependencies and communication.

We present \textbf{HeteroTask}, an end-to-end framework that connects kernel-wise multi-variant profiling, PEFT-based planning, host-buffer lifetime modeling, and distributed MPI execution.
A runtime-feedback loop uses Bayesian optimization (BO) to calibrate empirical overhead terms and regenerate plans.
The framework carries task, buffer, and placement information through these stages, connecting the scheduler's view of a workflow to the operations needed to execute it.
HeteroTask builds on established PEFT and BO methods.
Its contribution is their integration with multi-variant profiling, buffer modeling, and MPI execution, together with an evaluation of the integrated system.

Our evaluation covers five benchmark and workflow-derived DAGs on up to six nodes and 48 GPUs.
On two 8-GPU nodes, the selected HeteroTask executions have lower application-level completion times than the compared StarPU configurations across all five workloads.
The reductions range from 19.7\% to 72.1\% when the additional faster CED baseline is included.
Section~\ref{subsec:end2end} describes the conditions of that baseline.
The CED study also shows the value of exposing both CPU and GPU variants and of accounting for overheads in planning.
At the same time, the measurements distinguish three objectives that need not improve together: model agreement, task-window makespan, and application-level completion time.

The paper makes three contributions:
\begin{itemize}
\item \textbf{An integrated execution framework for fixed heterogeneous workflows.} HeteroTask connects per-variant profiles, communication estimates, and host-buffer metadata to placement and schedule-driven MPI execution.
\item \textbf{A runtime-feedback calibration mechanism.} Empirical corrections account for effects beyond isolated profiles.
For CED, the selected vector reduces task-window makespan by 25.0\% relative to the zero-overhead vector and by 3.5\% relative to the initial nonzero vector.
\item \textbf{A system-level evaluation across five workflows.} End-to-end timing, a CPU/GPU variant study, and fixed-work resource sweeps show where integrated planning and execution are effective.
They show that model agreement, task-window makespan, and application completion time respond differently to calibration and allocation.
\end{itemize}
Section~\ref{sec:Experiment} defines how executions were measured and selected, and Section~\ref{sec:limitations} states the scope of the evidence.
Profiling and calibration are preparation costs, so the value of plan reuse depends on the per-invocation savings and on how often the workflow runs.

\section{Related Work}\label{sec:Related_Work}
\begin{figure*}[t]
\centering
\includegraphics[pagebox=cropbox,clip,width=\textwidth]{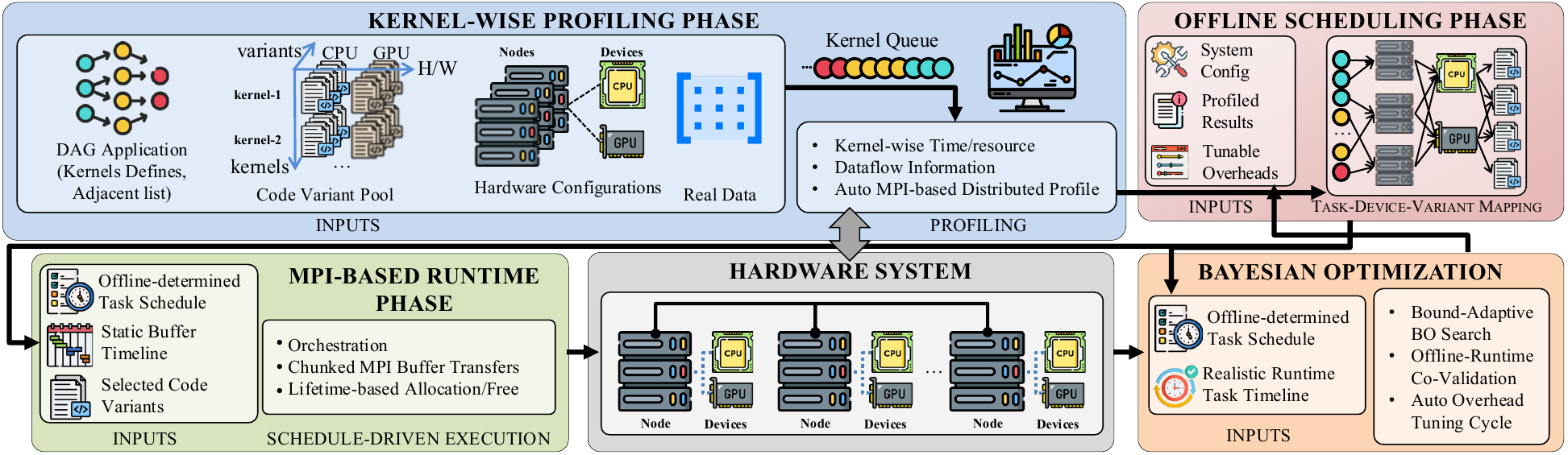}
\caption{HeteroTask's profile--schedule--execute pipeline on an exclusively allocated heterogeneous system. The runtime preserves the offline mapping and uses host-buffer metadata to construct dependency-driven allocation/release events. Plan--trace comparison feeds overhead calibration and regeneration of the offline plan.}
\label{fig:overview}
\end{figure*}

\subsection{Offline DAG Scheduling}
HEFT ranks tasks and then selects processors using estimated finish times~\cite{heft}.
PEFT adds optimistic downstream-cost lookahead~\cite{peft}.
Other heuristics explore placement consequences~\cite{l-heft}, constrained critical paths~\cite{ceft}, prediction-based prioritization~\cite{ippts}, or task duplication~\cite{hltsd}.
Genetic algorithms~\cite{genetic} and hybrid discrete particle-swarm optimization~\cite{hdpso} search a broader space of candidate schedules.
These methods provide different ways to construct an assignment from a cost model.
HeteroTask uses PEFT as its planning substrate and focuses on supplying that model with multi-variant profiles and runtime-calibrated costs, then realizing the resulting plan in a distributed executor.

\subsection{Task Runtimes and Variant Selection}
StarPU expresses heterogeneous tasks as multi-variant codelets and uses performance models to guide runtime decisions~\cite{starpu}. Taskflow, PaRSEC, and Legion provide related mechanisms for task-based execution, dependency management, and heterogeneous placement~\cite{taskflow,parsec,legion}. Such runtimes can react to observed readiness and resource availability. HeteroTask instead preserves an offline mapping under a fixed-workflow, exclusive-allocation contract. Comparative work on heterogeneous Cholesky factorization shows the value of static knowledge even under estimation error~\cite{agullo2016static}, while recent WORKS research adapts classic heuristics to online execution with partial graph knowledge~\cite{chamorro2025online}. These studies place HeteroTask within a broader spectrum of trade-offs between global planning and runtime adaptation.

Variant selection also arises below the distributed workflow level. MLCD selects code-version/device pairs using input characteristics~\cite{mlcd}, and FlexTensor explores tensor implementation schedules on heterogeneous hardware~\cite{flextensor}. HeteroTask coordinates the selection of available task variants with placement and communication across a complete DAG.

\subsection{Workflow Systems and Performance Calibration}
Pegasus manages distributed scientific workflows~\cite{pegasus}.
Ray supports distributed task and actor execution~\cite{ray}, and Dask.distributed schedules task graphs with data-locality awareness~\cite{dask}.
These systems address broader orchestration needs, while HeteroTask concentrates on kernel-wise placement and execution within an allocated CPU--GPU cluster.
Workflow performance-characterization research combines task provenance with I/O and operating-system observations~\cite{gueroudji2024workflow}, reinforcing the importance of measuring the whole application rather than its task region alone.

Gaussian-process BO provides a method for tuning expensive black-box objectives~\cite{snoek2012practical}, and constrained BO incorporates uncertain feasibility conditions~\cite{gelbart2014constraints}.
HeteroTask applies these principles to empirical overhead parameters: each candidate generates a plan whose execution supplies timing and alignment feedback.
The resulting loop connects the planning model to observed system behavior.
Our evaluation studies this integration with PEFT and the MPI runtime.

\section{HeteroTask Framework Design} \label{sec:HeteroTask_Design}

\subsection{Overview}

Fig.~\ref{fig:overview} shows the overall workflow of HeteroTask. Given a fixed DAG-form application and an exclusively allocated heterogeneous system, HeteroTask follows an end-to-end \emph{profile--schedule--execute} pipeline. The profiler first measures kernel-wise properties of each task variant on each device, including execution cost, resource metadata, and dataflow information. These measurements are then used by the offline scheduler, which decides task placement across servers and devices, code-variant selection, and data movement with communication estimates and host-buffer lifetime checks. The scheduling design space includes code variants for different backends, such as CPU/OpenMP, GPU/OpenMP, and CUDA.

The resulting schedule is executed by a lightweight C++/MPI runtime that preserves the offline assignments, manages buffers and transfers, and enforces dependencies and synchronization.
Runtime traces then feed a BO loop that adjusts overhead parameters and regenerates the plan.
Offline-Runtime Co-Validation in Fig.~\ref{fig:overview} is this plan--trace comparison: predicted and observed timings for a candidate plan are compared during calibration.
This feedback connects the planner's cost estimates to the executor's behavior.

\subsection{Kernel-wise Profiling Design}

HeteroTask uses kernel-wise profiling to feed the offline scheduler with device-specific execution data.
An MPI-based workflow automatically traverses all server nodes, profiles all hardware components, and merges the results into a single profile description,  giving the user a unified interface.

For each task variant on each device, the profiler logs latency and resource metadata.
CPU variants also capture thread configuration, while GPU variants are bound to explicit devices per server.
The profiled task input sizes are part of the scheduling contract and must match the later execution. Device-transfer and inter-server bandwidth measurements are stored in the system configuration and reused by the scheduler.
Beyond timing measurements, HeteroTask constructs task-level dataflow metadata that describes input/output buffers, dependency destinations, and buffer sizes.
Consequently, the profiling phase returns both execution costs and the graph structure needed for the planner.

\subsection{Offline Scheduling Design}

The offline scheduler takes per-variant profiling data, the application DAG, the target system configuration, and a tunable overhead vector $\theta$ (calibrated in Section~\ref{subsec:bo}) as inputs.
Represent the application DAG as $(V,\mathcal{D})$, where $V$ is the set of tasks and $\mathcal{D}$ is the set of directed data-dependency edges. For each task $i\in V$, let $K_i$ be its available code variants, and let $R$ be the set of CPU and GPU resources across all servers.
Let $a=\{(r_i,k_i)\}_{i\in V}$ denote a complete assignment, where $r_i\in R$ and $k_i\in K_i$, and let $\Omega$ be the assignment space defined by precedence, resource--variant compatibility, and communication requirements. The planning objective is
\begin{equation}
\begin{aligned}
\min_{a\in\Omega}\quad &C_{\max}(a;\theta),\\
C_{\max}(a;\theta)&=\max_{i\in V}F_i(a;\theta),
\end{aligned}
\label{eq:offline_objective}
\end{equation}
where $F_i(a;\theta)$ is the predicted finish time of task $i$ under assignment $a$ and overhead vector $\theta$, and $C_{\max}(a;\theta)$ is the resulting predicted makespan of the plan.

For a candidate placement $(i,r,k)$, communication time is the buffer size divided by the measured bandwidth of each segment on the transfer path, summed over the segments.
A transfer between two GPUs on one server passes through host memory.
A cross-server transfer adds a device-to-host copy at a GPU source and a host-to-device copy at a GPU destination around the network segment.
Runtime buffers reside in host memory, so an edge between consecutive tasks on the same GPU is also charged both copies.
HeteroTask applies overhead corrections separately to dependency readiness and task duration. Let $q$ denote the partial schedule and let $h_{ji}(\theta,q,r)$ collect the fixed setup, source-buffer, scheduling-progress, cross-device transition, and cross-server corrections associated with a predecessor edge. Then the dependency-ready time is
\begin{equation}
D_i(r;\theta,q)=\max_{j\in pred(i)}\{F_j+\tau_{ji}(r_j,r)+h_{ji}(\theta,q,r)\}.
\label{eq:edge_overhead}
\end{equation}
Here $\tau$ is the bandwidth-based transfer estimate, excluding the empirical corrections. The buffer term depends on the number of source-server output buffers whose consumers have not yet been assigned.
It adds $\alpha_b$ per buffer up to $b_0$ buffers and $\gamma_b\alpha_b$ per buffer beyond $b_0$.
The progress term is $\gamma_p\alpha_p$ times the fraction of tasks already scheduled.
These quantities describe the partial plan: the former is distinct from simultaneous live-buffer count, and the latter from elapsed execution time.

The availability search starts from $D_i$ and checks resource occupancy and provisional host-buffer intervals. Let $S_i$ be the returned start time and $T^{prof}_{i,r,k}$ the isolated profile. For an ordinary computational task, the candidate finish time is
\begin{equation}
F_i^{cand}=S_i+T^{prof}_{i,r,k}+u_{\mathrm{GPU}}(r;\theta)+u_{\mathrm{switch}}(r;\theta),
\label{eq:overhead_model}
\end{equation}
where $u_{\mathrm{GPU}}$ is zero for CPU placements and the configured GPU-task correction otherwise, while $u_{\mathrm{switch}}$ is the CPU- or GPU-transition correction. Virtual entry/exit tasks receive separate handling. This separation lets the model account for overheads that delay input readiness as well as those that extend execution on a worker.

The planner also maintains output-buffer lifetime intervals for each server. During candidate evaluation, it compares the total buffer volume over these intervals with host-memory capacity:
\begin{equation}
\sum_{b:\,s_b=s,\; l_b^s\le t<l_b^e}m_b\le M_s^{host},
\label{eq:memory_constraint}
\end{equation}
where $s_b$ is the buffer's server and $M_s^{host}$ is its host-memory capacity.
GPU kernel variants allocate and release their own device storage.
Thus, this server-level check is not a per-GPU live-residency constraint.
For every plan evaluated in this paper that has a retained planning log, the post-planning scan reports no host-capacity violation.
Equation~\eqref{eq:memory_constraint} is a planning check, not a certified bound on execution memory.

PEFT ranks tasks using an optimistic cost table (OCT) and chooses placements by combining candidate finish time with downstream lookahead.
HeteroTask evaluates resource--variant pairs using the profiled costs, empirical corrections, and host-buffer checks above.
It can additionally compare enabled CPU-only and GPU-only plans by predicted makespan.
Algorithm~\ref{alg:offline_scheduler} summarizes candidate evaluation for ordinary tasks.
Virtual entry and exit tasks are anchored to the master CPU.
The output consists of the selected mapping, predicted timing, and buffer metadata consumed by the runtime.
\begin{algorithm}[t]
\caption{Multi-variant candidate evaluation with PEFT.}
\label{alg:offline_scheduler}
\algrenewcommand\algorithmicindent{0.7em}
\small
\begin{algorithmic}[1]
\State \textbf{Input:} DAG, resources, variants, profiles, bandwidths, host capacities, $\theta$
\State \textbf{Output:} mapping, predicted times, provisional host-buffer timeline
\State Compute OCT and a precedence-respecting PEFT order
\State Initialize resource intervals, host-buffer intervals, and partial schedule $q$
\For{each ordinary task $i$ in order}
  \State $bestScore\gets\infty$
  \For{each compatible resource--variant pair $(r,k)$}
    \State $D\gets\max_{j\in pred(i)}\{F_j+\tau_{ji}+h_{ji}(\theta,q,r)\}$
    \State $T\gets T^{prof}_{i,r,k}+u_{\mathrm{GPU}}(r;\theta)$
    \State $S\gets$ \Call{BoundedAvailabilitySearch}{$D,T,r,q,M^{host}$}
    \State $F^{cand}\gets S+T+u_{\mathrm{switch}}(r;\theta)$
    \State $score\gets F^{cand}+OCT[i,r]$
    \If{$score<bestScore$}
      \State Record $(r,k,S,F^{cand})$ as the best candidate
      \State $bestScore\gets score$
    \EndIf
  \EndFor
  \State Commit candidate and update resource and host-buffer intervals
\EndFor
\State \Return mapping, times, and host-buffer metadata
\end{algorithmic}
\end{algorithm}

\subsection{Runtime Execution Design}

The HeteroTask runtime executes the offline schedule and associated memory timeline while handling dependency tracking, communication, and synchronization across MPI ranks.
MPI rank~0 maintains global state and dispatches tasks.
Each server contributes one CPU worker rank and one worker rank per selected GPU.
The CPU worker invokes the configured threaded variant internally.
This centralized design provides a simple schedule-preserving control path for the evaluated applications.
Dispatch follows three principles:

\begin{itemize}
\item \textbf{Dependency and Communication Orchestration:} Before dispatch, the master checks predecessor completion, buffer allocation, and input locality, then issues the required point-to-point transfers and prefetches.

\item \textbf{Schedule-Preserving Delay Propagation:} In the default schedule-timed mode, tasks wait for their planned start and dependencies.
Delays postpone dependents without changing placement.
The optional smart mode can release ready tasks earlier but does not remap them.

\item \textbf{Dependency-Driven Host-Buffer Management:} Producer and consumer metadata define when buffers must be available and can be released. Allocation and release events are ordered by these dependencies, so buffer lifetimes follow actual progress rather than predicted timestamps. Worker buffers reside in host memory, while GPU variants manage their own device allocations.
\end{itemize}

In the C++/MPI runtime, large point-to-point transfers are automatically split into multiple MPI messages and reassembled at the receiver, avoiding single-message limits while preserving the applications' logical buffer abstraction.

\subsection{Runtime-Feedback Calibration via Bayesian Optimization}\label{subsec:bo}

Real executions introduce launch, buffer-management, device-switch, and cross-server effects that isolated profiles do not capture. HeteroTask represents them with $\theta$ and calibrates this vector through the offline--runtime Bayesian-optimization (BO) loop in Fig.~\ref{fig:bo_tuning_flowchart}.

\begin{figure}[t]
\centering
\fbox{\parbox{.9\linewidth}{\textbf{Evaluate candidates}\\Overhead vector $\theta$ $\rightarrow$ offline plan $\rightarrow$ runtime trace $\rightarrow$ measured makespan and alignment.}}
\par\smallskip$\downarrow$\par\smallskip
\fbox{\parbox{.9\linewidth}{\textbf{Refine search}\\Fit surrogate models, update parameter bounds, and propose candidates under progressively tighter error thresholds, then repeat evaluation.}}
\par\smallskip$\downarrow$\par\smallskip
\fbox{\parbox{.9\linewidth}{\textbf{Select and reuse}\\Choose the fastest observed candidate, optionally requiring the final error thresholds. Save its parameters and plan for repeated execution.}}
\caption{Runtime-feedback calibration. Plan--trace errors guide search. The final selection rule determines whether model-alignment thresholds constrain the chosen plan.}
\label{fig:bo_tuning_flowchart}
\end{figure}

For each candidate $\theta$, let $a_\theta\in\Omega$ be the assignment produced by the offline scheduler and executed by the runtime. Define $C^{off}(\theta)=C_{\max}(a_\theta;\theta)$, and let $C^{rt}(\theta)$ be the measured runtime makespan of that assignment. The relative makespan error $E^{mk}(\theta)$ is
\begin{equation}
E^{mk}(\theta)=\frac{|C^{rt}(\theta)-C^{off}(\theta)|}{C^{off}(\theta)}.
\label{eq:makespan_error}
\end{equation}
For task-level alignment, let $V'\subseteq V$ be the tasks observed in both traces, with $|V'|$ denoting their number. For each $i\in V'$, let $(r_i,k_i)$ be its selected resource--variant pair in $a_\theta$. Then $S_i^{off}(\theta)=S_i(r_i,k_i;\theta)$ and $F_i^{off}(\theta)=F_i(a_\theta;\theta)$ are its predicted start and finish times, while $S_i^{rt}(\theta)$ and $F_i^{rt}(\theta)$ are their measured runtime counterparts. The normalized average start- and finish-time deviations, $\bar{E}^{s}(\theta)$ and $\bar{E}^{f}(\theta)$, are
\begin{align}
\bar{E}^{s}(\theta)
&=\frac{1}{|V'|}\sum_{i\in V'}
\frac{|S_i^{rt}(\theta)-S_i^{off}(\theta)|}{C^{rt}(\theta)}, \nonumber\\
\bar{E}^{f}(\theta)
&=\frac{1}{|V'|}\sum_{i\in V'}
\frac{|F_i^{rt}(\theta)-F_i^{off}(\theta)|}{C^{rt}(\theta)}.
\label{eq:task_alignment_error}
\end{align}
We use weights $(w_M,w_s,w_f)=(3,1,1)$ to give greater weight to overall makespan agreement.
The start- and finish-time terms penalize a vector whose predicted makespan matches the execution while individual tasks start or finish at different times.
Unlike the makespan-only error above, the makespan term inside alignment is normalized by the observed runtime makespan. The alignment error is
\begin{equation}
E^{align}(\theta)=3\frac{|C^{rt}(\theta)-C^{off}(\theta)|}{C^{rt}(\theta)}
+\bar{E}^{s}(\theta)+\bar{E}^{f}(\theta).
\label{eq:alignment_error}
\end{equation}
The search models task-window makespan and both errors with Gaussian-process regressors using Mat\'ern kernels. It proposes candidates using expected improvement over the fastest observation that meets the current thresholds, or the fastest overall if none does, weighted by estimated threshold-feasibility probabilities, following established GP-based and constrained BO principles~\cite{snoek2012practical,gelbart2014constraints}.
The feasibility probability is the product of the probabilities that each error meets its threshold.
Each observation requires generating and executing a plan, making sample efficiency an important consideration.

Search begins with loose error thresholds and expands parameter bounds when too few candidates satisfy them. Subsequent rounds narrow the bounds around promising groups and tighten the thresholds. This process combines exploration with progressively stronger requirements on model agreement.
Each round refits the regressors to all earlier observations and reclassifies them under its own thresholds, so tightening the thresholds reuses earlier executions rather than discarding them.
An execution marked as failed or timed out enters the dataset with penalty values and counts as infeasible.
Table~\ref{tab:calibration_parameters} lists the coefficients of $\theta$ and the initial bounds from which this refinement starts.
The coefficients jointly correct the planning model rather than represent independently measured hardware latencies.
For CED, the search evaluates 62 vectors in five rounds.
The first round contains 22 vectors, including the initial nonzero vector $\theta_0$ and a zero-overhead reference, and each subsequent round evaluates ten.
The thresholds $(\epsilon_M,\epsilon_A)$ on $E^{mk}$ and $E^{align}$ are $(1.0,2.0)$, $(0.7,1.2)$, $(0.4,0.9)$, $(0.2,0.7)$, and $(0.1,0.5)$ in the five rounds.

Final selection depends on the objective of plan reuse.
One option selects the fastest observed candidate satisfying both final error thresholds.
Another selects the fastest observed execution regardless of alignment.
The evaluated CED and two-node 3MM searches use the second option.
CED's selection also meets the error thresholds, while 3MM's does not (Section~\ref{sec:Results}).
These thresholds concern model agreement, not execution correctness or memory capacity.
Algorithm~\ref{alg:bo_overhead_tuning} makes the final selection rule explicit.

\begin{algorithm}[t]
\caption{Runtime-feedback calibration with BO.}
\label{alg:bo_overhead_tuning}
\algrenewcommand\algorithmicindent{0.8em}
\small
\begin{algorithmic}[1]
\State \textbf{Input:} parameter bounds $B$, seed parameters $\theta_0$, threshold schedule $\{(\epsilon_M,\epsilon_A)\}$
\State \textbf{Output:} selected overhead parameters $\theta^\star$
\State $D \gets \emptyset$
\State $\Theta \gets$ \Call{InitialSamples}{$B,\theta_0,\mathbf{0}$} \Comment{initial vector, zero reference, and samples}
\For{each threshold pair $(\epsilon_M,\epsilon_A)$ in the schedule}
  \While{true}
    \For{each $\theta \in \Theta$}
      \State $(S_\theta,L_\theta) \gets$ \Call{OfflineSchedule}{$\theta$}
      \State $\mathcal{T}_\theta \gets$ \Call{RuntimeExecute}{$S_\theta,L_\theta$}
      \State $M^{rt}_\theta \gets$ \Call{Makespan}{$\mathcal{T}_\theta$}
      \State $E^{mk}_\theta \gets$ \Call{MakespanError}{$S_\theta,\mathcal{T}_\theta$}
      \State $E^{align}_\theta \gets$ \Call{AlignmentError}{$S_\theta,\mathcal{T}_\theta$}
      \State $D \gets D \cup \{(\theta,M^{rt}_\theta,E^{mk}_\theta,E^{align}_\theta)\}$
    \EndFor
    \State $F \gets \{x \in D \mid x.E^{mk} \le \epsilon_M \land x.E^{align} \le \epsilon_A\}$
    \If{$|F|$ is too small}
      \State $B \gets$ \Call{ExpandBounds}{$B$}
      \State $\Theta \gets$ \Call{SuggestBO}{$D,B,\text{global}$}
    \Else
      \State $B \gets$ \Call{RefineBounds}{$B,\text{BestGroup}(F)$}
      \State $\Theta \gets$ \Call{SuggestBO}{$D,B,\text{local}$}
      \State \textbf{break}
    \EndIf
  \EndWhile
\EndFor
\State $\theta^\star \gets$ \Call{SelectFinalTheta}{$D,\text{final-choice}$} \Comment{fastest threshold-feasible or fastest observed}
\State Save $\theta^\star$ and its plan for subsequent executions
\State \Return $\theta^\star$
\end{algorithmic}
\end{algorithm}

The selected vector and plan can be reused while the workflow, input shapes, allocation, and execution conditions remain sufficiently stable. This shifts profiling, search, and candidate execution into a preparation phase. Section~\ref{sec:Results} considers the cost of this preparation alongside per-invocation performance.

\subsection{Application Integration Example}
To integrate an application, the developer supplies task variants, buffer declarations, and wrappers that connect inputs and outputs to each kernel.
For CED, a Gaussian-filter task consumes an image buffer and produces an intermediate image for downstream stages.
Profiling records the variants' costs, buffer sizes, and consumers.
The planner selects a server, device, and variant, and the runtime invokes that variant with its buffers while handling dependency events and inter-rank movement.
The same task and dataflow description supports profiling, placement, and execution.
Application kernels remain separate from distributed orchestration.

\section{Experimental Setup}\label{sec:Experiment}
\subsection{Platform and Workloads}
\begin{figure*}[t]
    \centering
    \begin{minipage}[t]{0.22\textwidth}
        \centering
        \includegraphics[height=2.45cm]{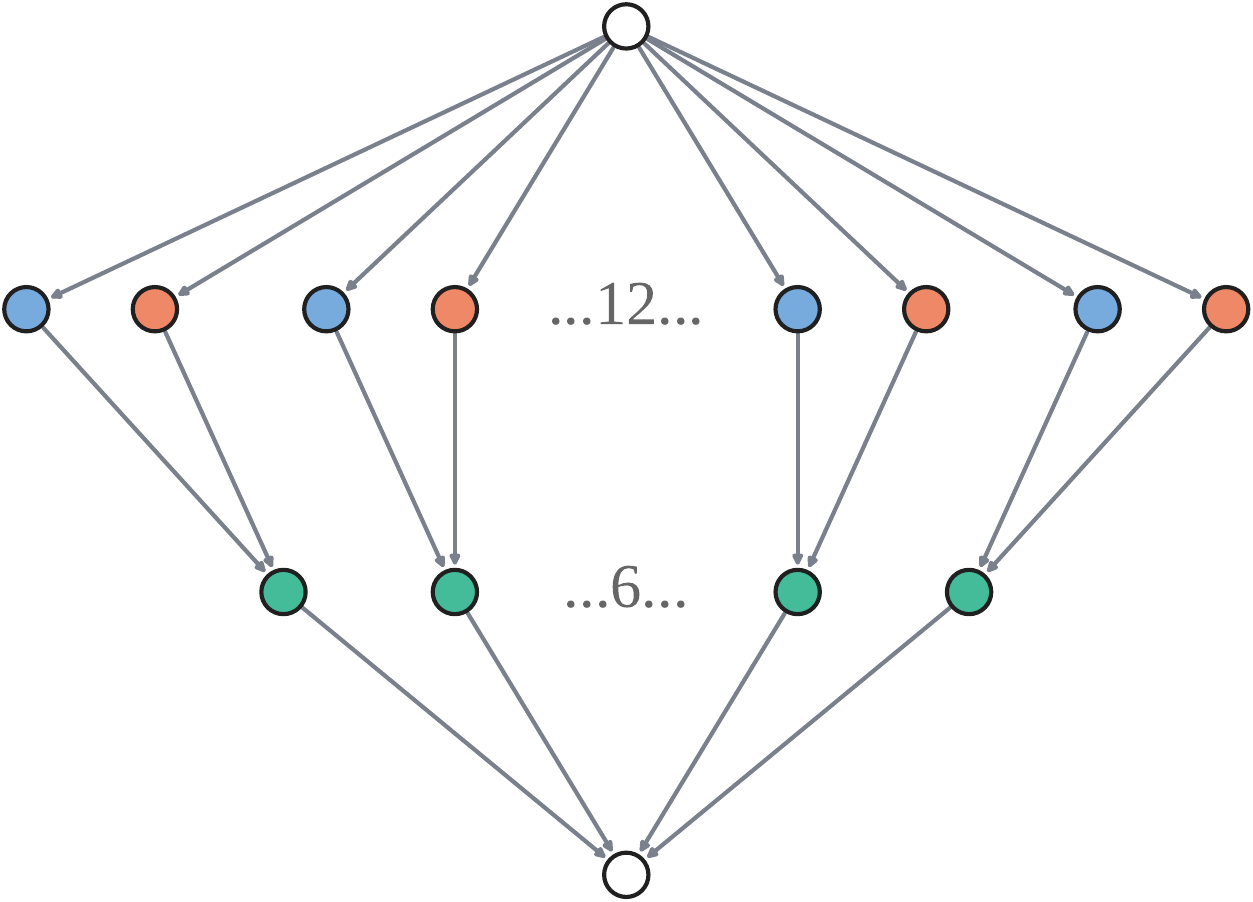}
        \vspace{2pt}

        {\small (a) 3MM / SBF}%
    \end{minipage}
    \hspace{0.01\textwidth}
    \begin{minipage}[t]{0.22\textwidth}
        \centering
        \includegraphics[height=2.45cm]{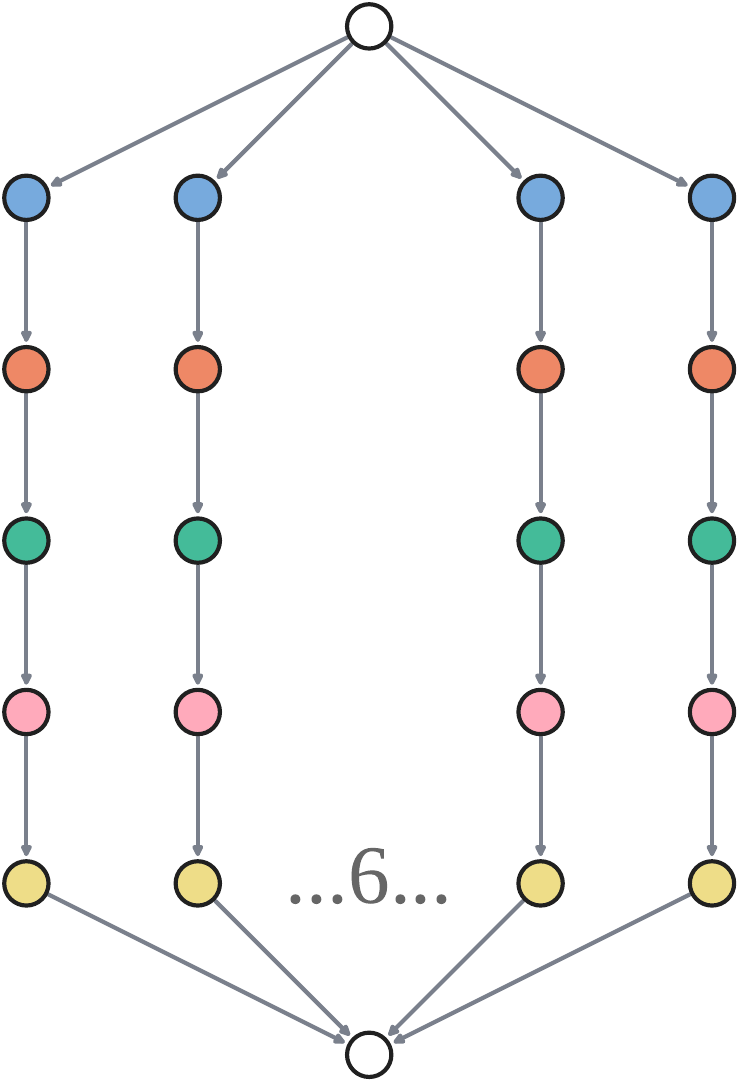}
        \vspace{2pt}

        {\small (b) CED}%
    \end{minipage}
    \hspace{0.01\textwidth}
    \begin{minipage}[t]{0.20\textwidth}
        \centering
        \includegraphics[height=2.45cm]{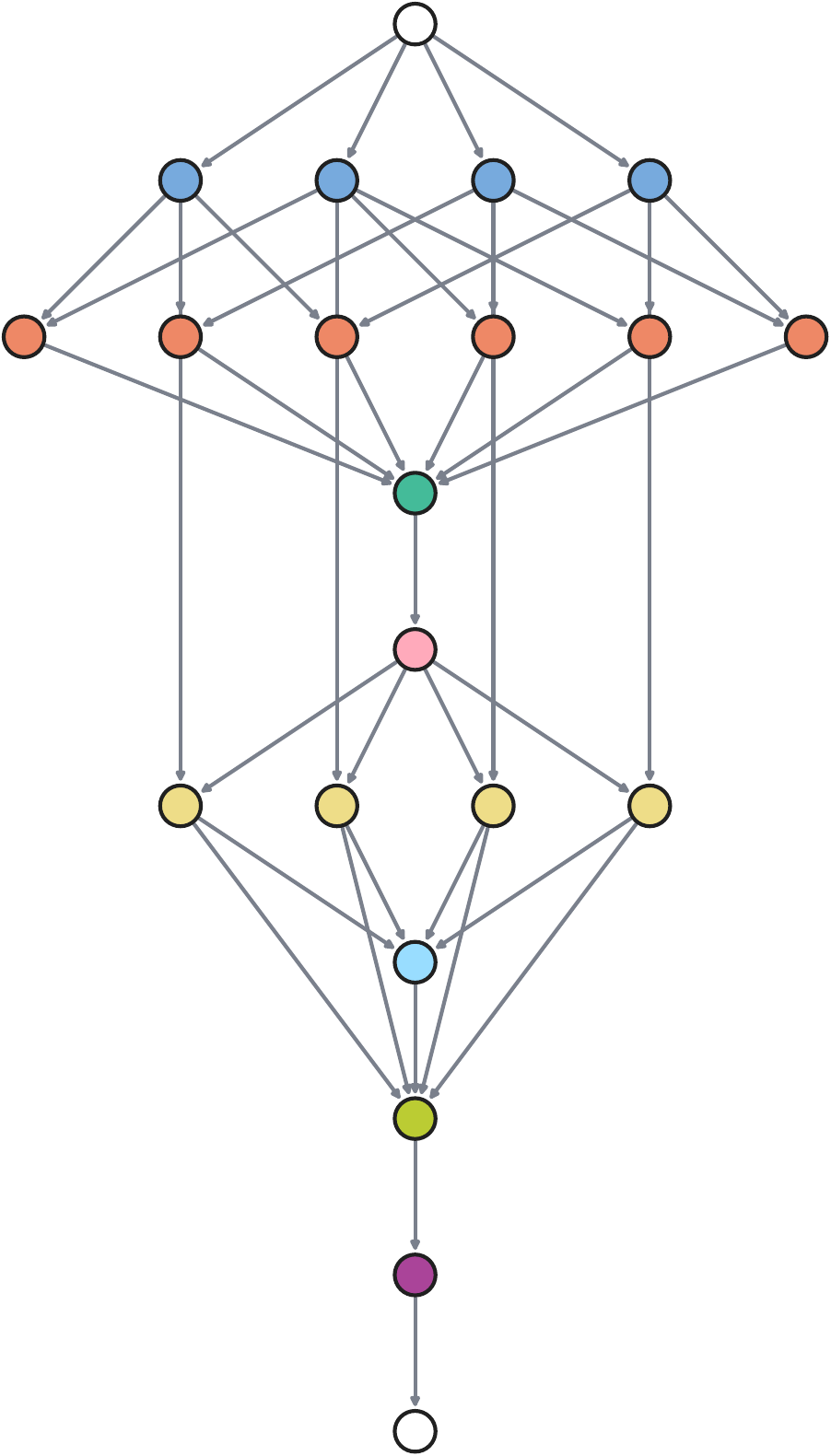}
        \vspace{2pt}

        {\small (c) Montage}%
    \end{minipage}
    \begin{minipage}[t]{0.31\textwidth}
        \centering
        \includegraphics[height=2.45cm]{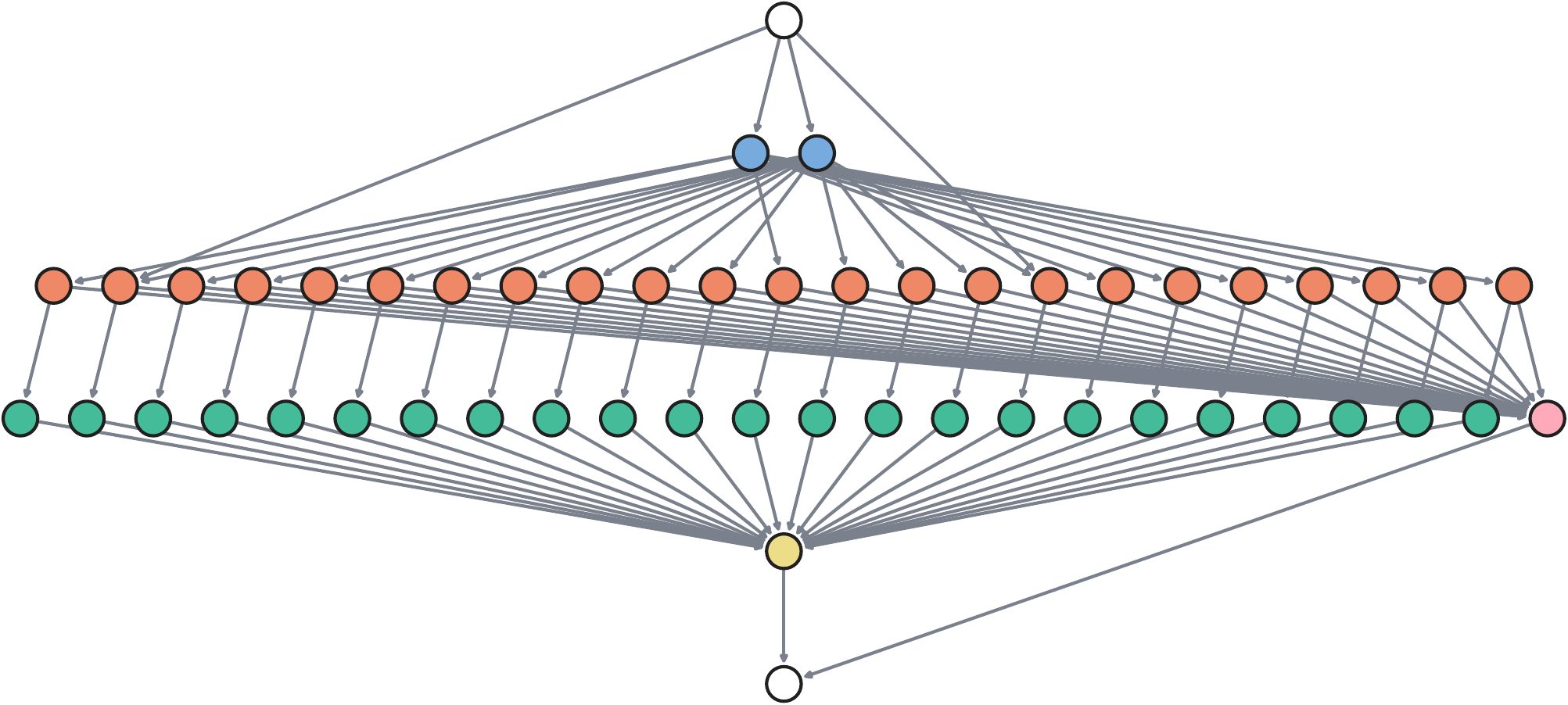}
        \vspace{2pt}

        {\small (d) CyberShake}%
    \end{minipage}
    \caption{DAGs of the five C++ applications used in evaluation. 3MM, CED, and SBF are transformed into 10-instance batched DAGs via automatic pipeline unrolling. 3MM and SBF share the same unrolled topology, so it is shown once in (a). Montage and CyberShake use Pegasus-generated DAGs with irregular dependencies.}
    \label{fig:cpp_application_dags}
    
\end{figure*}
\begin{table*}[t]\centering\scriptsize
\caption{Workload characteristics. Declared output-buffer volume counts each task's first available profile once. It is not the peak resident memory of an execution.}\label{tab:app_characterization}
\begin{tabular}{lccccc}\toprule
Property & 3MM & CED & SBF & Montage & CyberShake\\
\midrule
Domain & Dense LA & Image proc. & Image proc. & Sci. workflow & Sci. workflow\\
Tasks & 32 & 52 & 32 & 21 & 52\\
Edges & 50 & 60 & 50 & 42 & 98\\
Workload setting & 10-way batch & 10-image batch & 10-image batch & 1 workflow inst. & 1 workflow inst.\\
Declared output-buffer volume & 4.25 GiB & 68.16 GiB & 81.72 GiB & 949.31 MiB & 1.30 GiB\\
Largest buffer & 128.00 MiB & 2.71 GiB & 2.71 GiB & 98.88 MiB & 329.76 MiB\\
\bottomrule\end{tabular}\end{table*}

We evaluate HeteroTask on allocations drawn from six nodes, each configured with an AMD EPYC 7763 64-core CPU, 1~TB of host memory, and eight NVIDIA A100-SXM4 GPUs with 80~GB each.
Allocations range from one node with one GPU to six nodes with eight GPUs per node.
The node type is uniform, so the evaluated heterogeneity is between CPU and GPU resources.
The two-node CED execution uses 19 MPI ranks: a master and one CPU worker plus eight GPU workers per node.
The inspected software snapshot specifies Clang 18.1.3, NVIDIA HPC SDK 26.1, CUDA 13.1, and a StarPU 1.4 interface.
These versions describe that snapshot rather than a recovered binary manifest for each run.

CED, Sobel Filter (SBF), and 3MM use ten pipeline instances unrolled into a single DAG~\cite{tian2025heterobench}.
Montage and CyberShake retain Pegasus-generated dependency structures with workflow-derived operators~\cite{pegasus}.
They do not execute the complete production applications through Pegasus.
The five DAGs contain 21--52 tasks, including virtual entry and exit tasks, and cover regular pipelines as well as irregular dependency structures.
We size the DAGs so that repeated profiling, baseline tuning, and BO calibration fit within the cluster's fixed job wall time, because a larger DAG leaves less time per kernel and makes kernel measurements more sensitive to noise.
Table~\ref{tab:app_characterization} reports their characteristics.
Declared output-buffer volume describes the graph's data objects, not live memory or peak GPU usage.

\subsection{StarPU Baselines}
The distributed StarPU implementations use the same underlying application kernels with StarPU-specific data-handle wrappers. We compare modes exposing CPU/OpenMP and CUDA codelets with modes restricted to CUDA codelets. The experiment protocol explores \texttt{dmda}, \texttt{eager}, and \texttt{heft}, with five warm-up runs followed by 8--15 formal runs and a stability check over the trailing five measurements. Policy names denote recorded configurations. Different names alone do not imply different scheduling algorithms.

CED and SBF use explicit data partitioning and bounded pipelining in StarPU. These choices make the large-image workloads executable but change task granularity and data placement. HeteroTask's benchmark-style GPU variants include host--device transfers inside tasks, whereas StarPU can stage and reuse data outside codelet bodies. The application-level comparison therefore combines the effects of placement, data management, and runtime orchestration.

\subsection{Measurements and Configuration Selection}\label{subsec:measurement}
Application end-to-end (E2E) time spans initialization and input setup through completion synchronization and output recovery. The task window spans the first scheduled task start to the last task completion. Start/setup and end/recovery are derived from each implementation's trace markers and represent broad phases, not matched microcomponents. Profiling, offline planning, and BO calibration are excluded from invocation E2E and discussed separately as preparation costs.

The available results retain selected execution traces rather than the complete repeated-run series.
We therefore report point measurements, without treating plotted bars as verified medians or adding confidence intervals.
The CED result selected by HeteroTask is also an observation used during calibration, and independent post-selection repeats are unavailable.
BO evaluations with different overhead vectors are different configurations, not repetitions of one configuration.

Figure~\ref{fig:two_node_e2e} compares the selected CPU+GPU and CUDA-only StarPU executions with HeteroTask on two nodes. An additional CED execution has lower StarPU E2E and is shown to assess sensitivity to baseline selection. For the resource sweep, each allocation uses the lowest available StarPU E2E and the selected HeteroTask invocation. Full input, build, and correctness equivalence is not documented for every configuration, so these comparisons characterize the measured implementations rather than a best achievable baseline.

We cross-check timing summaries against logs, verify phase sums, and recompute calibration errors from matched task traces.
Reductions are calculated as $(T_{\mathrm{ref}}-T_{\mathrm{HT}})/T_{\mathrm{ref}}$, with the reference given for each comparison.
All results are recomputed offline from the retained execution records, with no new application runs.
Source code and supporting experiment materials have not yet been publicly released.

\section{Results and Discussion}\label{sec:Results}
\subsection{Calibration of Planning and Execution}
\begin{figure*}[t]
\centering
\begin{subfigure}[t]{.32\textwidth}
\centering
\includegraphics[width=\linewidth]{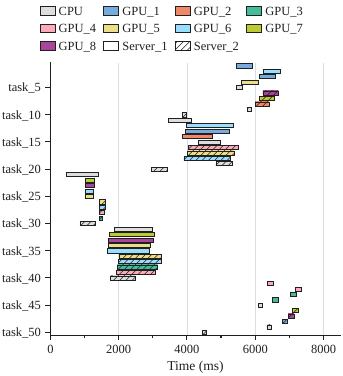}
\caption{Calibrated plan: 8.503 s}
\end{subfigure}\hfill
\begin{subfigure}[t]{.32\textwidth}
\centering
\includegraphics[width=\linewidth]{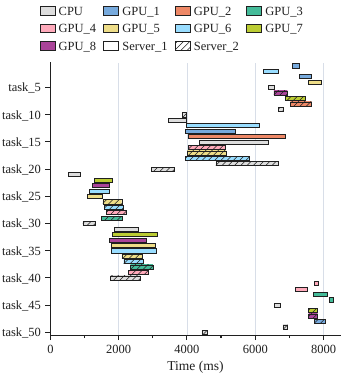}
\caption{CPU+GPU execution: 8.504 s}
\end{subfigure}\hfill
\begin{subfigure}[t]{.32\textwidth}
\centering
\includegraphics[width=\linewidth]{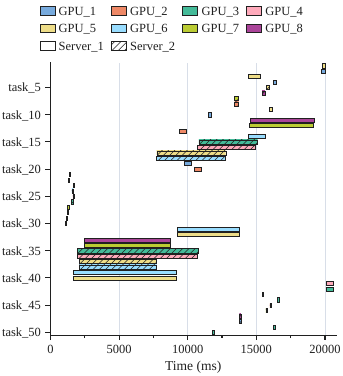}
\caption{GPU-only execution: 20.703 s}
\end{subfigure}
\caption{CED planning and variant-selection study on two 8-GPU nodes. The calibrated plan and CPU+GPU execution use the selected overhead vector, and the GPU-only execution uses the zero-overhead vector. Virtual entry/exit tasks are omitted from the visible rows but retained in the reported total spans. Rows use task-name ordering, not topological IDs. The CPU+GPU and GPU-only configurations differ in active computational resources.}
\label{fig:ablation_gantt}
\end{figure*}

The CED study connects the planning--execution gap in Fig.~\ref{fig:intro_gantt_gap} to the calibration mechanism.
With zero overhead corrections, the plan predicts 4.584~s and executes in 11.334~s.
After five search rounds evaluating 62 parameter vectors, the selected plan predicts 8.50346~s and executes in 8.50396~s.
Figure~\ref{fig:ablation_gantt} shows the resulting plan and execution timelines.
In Figs.~\ref{fig:intro_gantt_gap} and~\ref{fig:ablation_gantt}, the last visible computational task can end before the reported span, which includes the two omitted virtual tasks.
The alignment errors in Table~\ref{tab:ced_calibration} cover all 52 CED tasks, including these two virtual tasks.
\begin{table}[t]\centering\small
\caption{CED configurations within the calibration study. Times are task-window measurements.}\label{tab:ced_calibration}
\begin{tabular}{lrr}\toprule
Configuration & Task window (s) & Alignment error\\
\midrule
Zero overhead & 11.334 & 2.293\\
Initial nonzero & 8.814 & 0.108\\
Selected & 8.504 & 0.111\\
\bottomrule\end{tabular}\end{table}

Relative to the zero-overhead configuration, the selected execution reduces task-window makespan by 25.0\% and alignment error by 95.2\%.
The makespan error drops from 147.24\% to 0.00592\%.
The planner predicts a longer execution after calibration because it accounts for additional costs, yet the resulting plan executes faster.
This case illustrates why minimizing a prediction based only on isolated kernel costs can lead to a poor runtime result.

Table~\ref{tab:ced_calibration} also includes the initial nonzero vector, which already executes in 8.814~s.
The final selection improves on this reference by 3.5\%, while its alignment error is slightly larger.
The 25.0\% difference measures the benefit relative to the zero-overhead plan, whereas 3.5\% measures the gain beyond the initial vector.
The initial vector's earlier preparation history is unknown.
All three values are search observations.
The near-zero final residual describes agreement on the selected execution, not prediction accuracy on unseen runs.

\subsection{Calibration Search Trajectory}
\begin{table*}[t]\centering\small
\caption{Empirical overhead parameters: initial search ranges and selected CED values. Symbols name model coefficients, not independently measured hardware latencies.}\label{tab:calibration_parameters}
\begin{tabular}{lllr r}\toprule
Symbol & Model role & Unit & Initial range & CED selected\\
\midrule
$\alpha_0$ & Fixed dependency correction & ms & [0,1000] & 155.0\\
$b_0$ & Buffer-count threshold & count & [0,20] & 2.0\\
$\alpha_b$ & Per-buffer correction & ms/buffer & [0,200] & 4.0\\
$\gamma_b$ & Buffer-cost growth factor & dimensionless & [0,10] & 1.2\\
$\alpha_g$ & GPU-task correction & ms & [0,500] & 14.9\\
$\gamma_p$ & Scheduling-progress multiplier & dimensionless & [0,5] & 0.0\\
$\alpha_p$ & Progress scale before multiplier & ms & [0,10000] & 1038.4\\
$\sigma_c$ & CPU-task transition correction & ms & [0,5000] & 288.9\\
$\sigma_g$ & GPU-task transition correction & ms & [0,5000] & 172.4\\
$\sigma_{cg}$ & Cross-device edge correction & ms & [0,5000] & 641.1\\
$\alpha_n$ & Cross-server data correction & ms/GiB & [0,2000] & 231.0\\
\bottomrule\end{tabular}\end{table*}

\begin{figure*}[t]
\centering
\includegraphics[width=\linewidth]{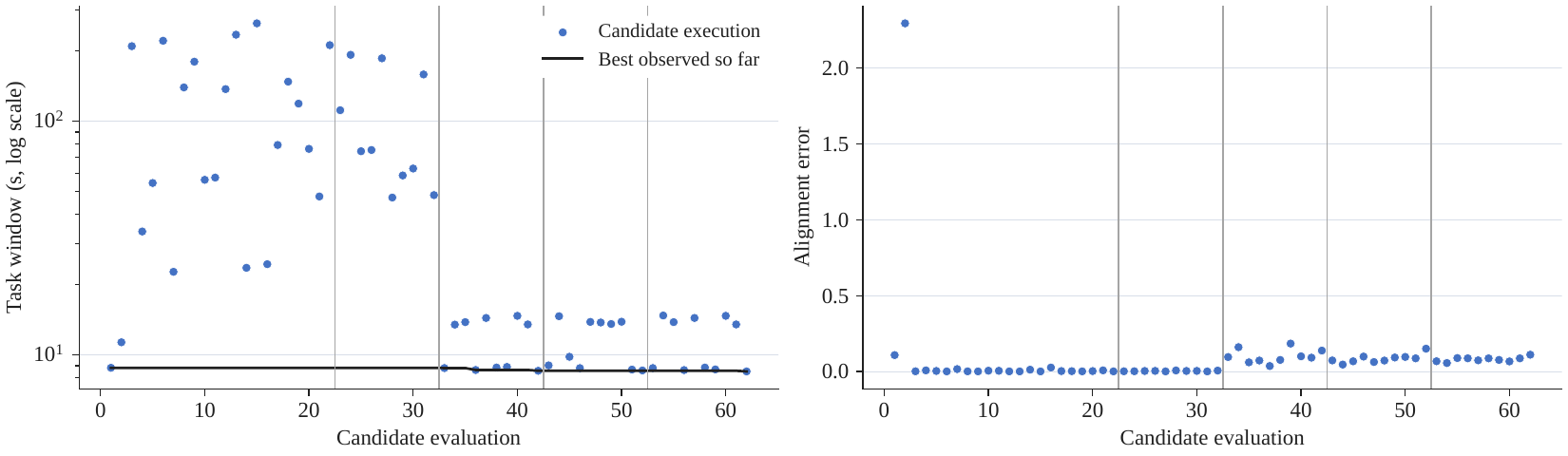}
\caption{CED calibration over 62 candidate evaluations. Left: task-window makespan and running minimum. Right: alignment error. Vertical lines separate search rounds. The initial nonzero vector already has low execution time, while some slower candidates have small alignment errors.}
\label{fig:ced_search}
\end{figure*}
Table~\ref{tab:calibration_parameters} gives the coefficient values of the selected CED vector.
In this vector, the zero progress multiplier disables its progress-dependent term.
Figure~\ref{fig:ced_search} shows the full search trajectory and the distinct behavior of execution time and alignment during tuning.
The running minimum improves modestly, from 8.814~s for the initial nonzero vector to 8.560~s in the third round and 8.504~s in the fifth, while individual task windows reach 262.610~s.
Every second-round candidate takes at least 47.178~s, whereas all 30 candidates in the last three rounds finish within 8.504--14.758~s.
All 61 candidates other than the zero-overhead reference meet the final error thresholds, so these thresholds alone do not identify the fastest plan.
This separation motivates evaluating measured speed alongside model agreement rather than using alignment alone to select a plan.

\subsection{Value of Heterogeneous Variant Selection}
Restricting all computational CED tasks to GPU variants under the zero-overhead vector yields a task window of 20.703~s.
Without calibration, enabling CPU variants on the same two-node allocation reduces the task window to 11.334~s, a 45.3\% reduction.
The calibrated CPU+GPU execution completes in 8.504~s and assigns to CPUs 14 computational tasks from all five CED stages (Gaussian filter 2, gradient 2, edge thinning 4, double thresholding 3, hysteresis 3), excluding the two virtual tasks.
Relative to the uncalibrated GPU-only execution, its 58.9\% reduction combines variant selection and calibration.
The timelines in Fig.~\ref{fig:ablation_gantt} show how these assignments distribute work beyond the GPU workers and permit additional overlap.

This comparison demonstrates the value of exposing the full heterogeneous allocation, including its CPU cores as additional compute resources, to the planner for CED.

\subsection{End-to-End Workflow Performance}\label{subsec:end2end}
\begin{figure*}[t]
\centering
\includegraphics[width=\textwidth]{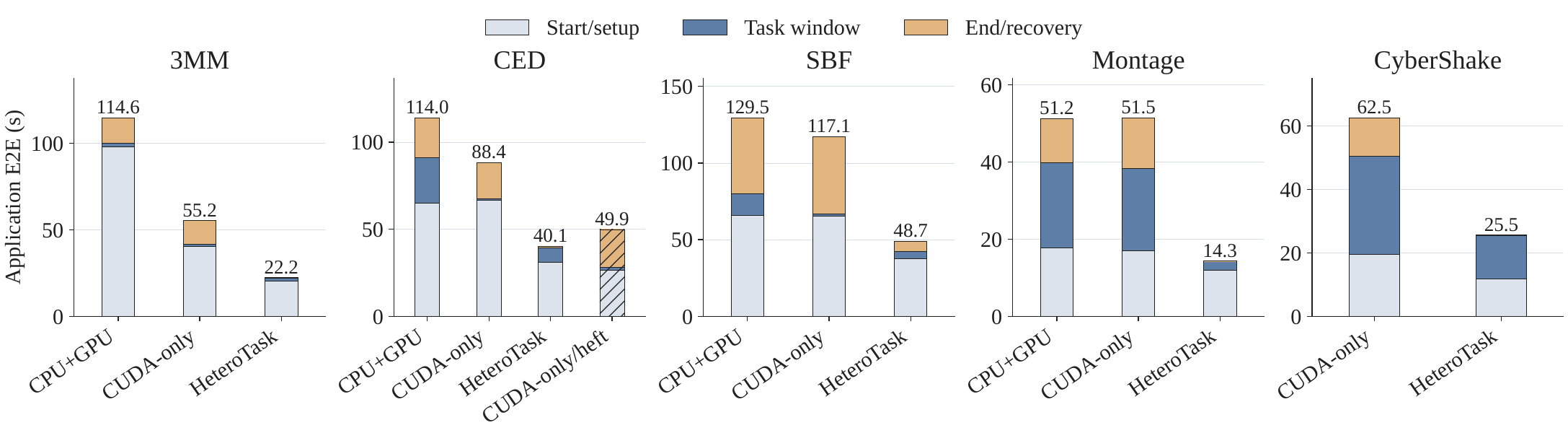}
\caption{Application-level timing on two 8-GPU nodes. CPU+GPU and CUDA-only denote StarPU modes. Start/setup, task-window, and end/recovery boundaries follow each implementation's markers. The hatched CED bar is the additional CUDA-only/\texttt{heft} execution. The CyberShake CPU+GPU result is unavailable and omitted.}
\label{fig:two_node_e2e}
\end{figure*}

On two 8-GPU nodes, the selected HeteroTask invocations have lower E2E than the compared StarPU configurations for all five workloads.
Of the two selected StarPU modes, CUDA-only is faster for 3MM, CED, and SBF, whereas CPU+GPU is faster for Montage by only 0.4\% and CyberShake has only a CUDA-only result.
Against the faster selected StarPU mode, the reductions are 59.7\% for 3MM, 54.7\% for CED, 58.4\% for SBF, 72.1\% for Montage, and 59.1\% for CyberShake.
Including the additional faster CED baseline changes CED's reduction to 19.7\%, and the resulting five-workload range is 19.7--72.1\%.
SBF illustrates variant selection: the planner chooses its all-CPU plan, which runs on one node.

The phase breakdown explains why application-level timing matters. For 3MM, CED, and SBF, a compared StarPU configuration has a shorter task window but a longer E2E than HeteroTask.
For Montage and CyberShake, HeteroTask also has the shorter task window: 2.290~s against 21.219~s and 22.110~s for the two StarPU modes, and 13.644~s against 30.807~s.
Data movement and management occur at different points in the two implementations. HeteroTask's selected CED invocation, for example, comprises 30.881~s of start/setup, 8.504~s of task execution, and 0.666~s of end/recovery. Its task window is only part of the total cost.
Start/setup is the largest HeteroTask phase for four of the five workloads, at 77.1--92.1\% of E2E, whereas CyberShake's task window exceeds its 11.875~s start/setup.
The E2E advantage thus reflects integration across planning, data handling, and execution, not the scheduling heuristic alone.

The two CED StarPU configurations make the selection issue concrete. The selected CUDA-only/\texttt{dmda} execution has a 0.926~s task window and 88.438~s E2E. The additional CUDA-only/\texttt{heft} execution has a longer task window of 1.643~s but a shorter E2E of 49.879~s. Both contain 800 partitioned task records for ten images. Figure~\ref{fig:two_node_e2e} shows the phases of both executions. This supports using application E2E to choose a runtime configuration and prevents the shorter task window from being mistaken for the stronger baseline.

\subsection{Resource Allocation at Fixed Workload Size}
\begin{figure*}[t]
\centering
\includegraphics[width=\textwidth]{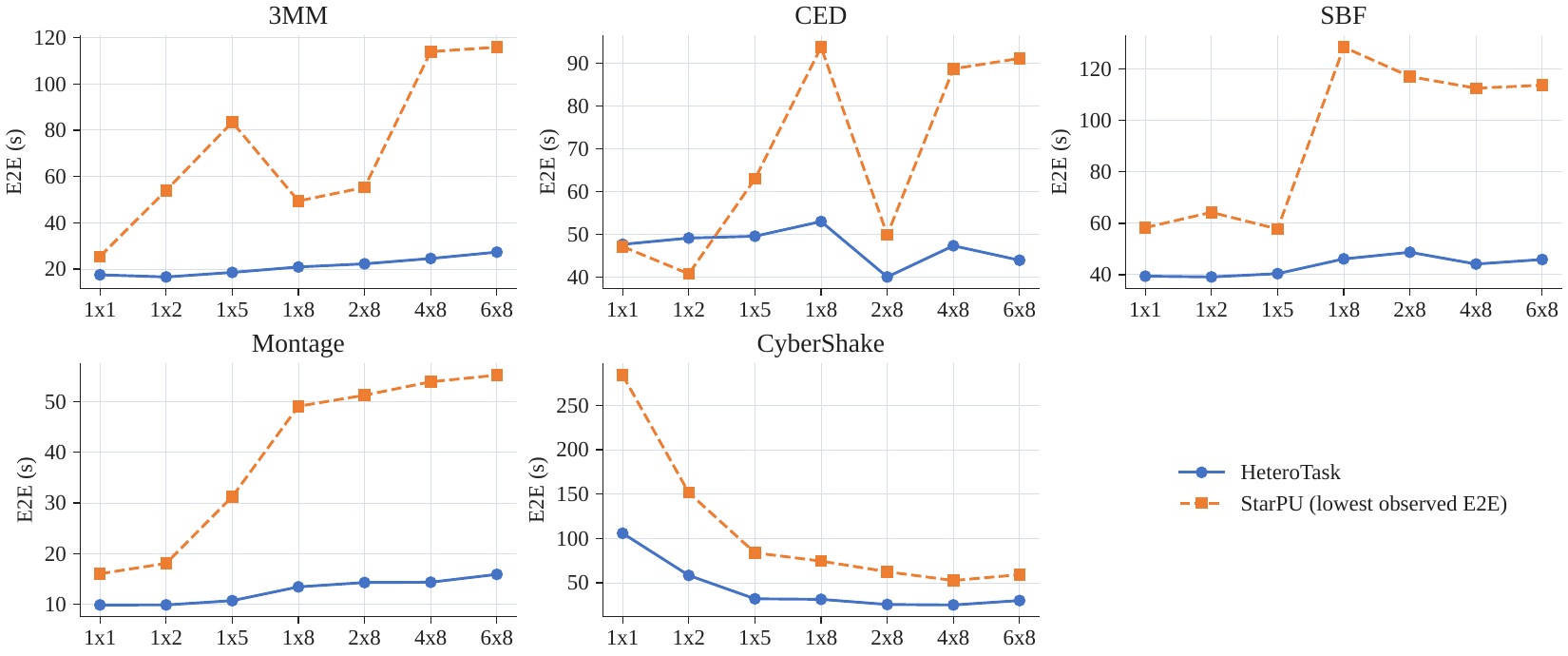}
\caption{Application E2E across resource allocations at fixed workload size. Labels denote nodes $\times$ GPUs per node. Each point uses the selected HeteroTask execution and the lowest available StarPU E2E.}
\label{fig:all_resource_records}
\end{figure*}
Figure~\ref{fig:all_resource_records} extends the comparison to all five workloads over allocations up to six nodes and 48 GPUs.
Against the lowest available StarPU E2E at each allocation, HeteroTask has the lower measured E2E at 33 of the 35 plotted application--allocation points, including every plotted allocation for 3MM, SBF, Montage, and CyberShake.
CED is more sensitive to allocation and baseline selection, and both exceptions are CED on one node: with one GPU, the times are 47.670~s for HeteroTask and 47.157~s for StarPU, and StarPU is also faster with two GPUs.
Across the plotted allocations, the lowest available StarPU CED E2E ranges from 40.8~s at 1$\times$2 to 93.7~s at 1$\times$8.
HeteroTask reduces E2E by 43.5--72.6\% across the five workloads at 1$\times$8 and by 49.3--76.4\% at 6$\times$8.
CyberShake's reduction ranges from 49.3\% at 6$\times$8 to 62.7\% at 1$\times$1.
At each allocation, the planner keeps the plan with the shortest predicted makespan among the enabled PEFT, all-CPU, and all-GPU plans.
It keeps the PEFT plan except for SBF at 1$\times$1, 1$\times$2, 2$\times$8, and 4$\times$8, where it keeps the all-CPU plan, and for 3MM at 1$\times$1, where it keeps the all-GPU plan.

Adding resources does not uniformly shorten these fixed DAGs.
HeteroTask's CED E2E is 40.1~s on two 8-GPU nodes, 47.3~s on four, and 44.0~s on six.
The six-node CED plan places all tasks on one node.
Its task window is 11.0~s, compared with 8.5~s on two nodes and 12.1~s on four.
CyberShake benefits more from additional resources, although its largest allocation is not the fastest either.
Its HeteroTask E2E is 105.8~s at 1$\times$1, 25.0~s at 4$\times$8, and 30.0~s at 6$\times$8.

These curves characterize resource choices for a fixed application, rather than weak scaling or growing-DAG scalability. Each allocation uses its own selected configuration. For six-node SBF, Montage, and CyberShake, the execution traces cannot be linked unambiguously to the final calibration selections. We retain the measured application times but do not attribute those points to particular calibrated vectors.

\subsection{Calibration Trade-offs and Plan Reuse}
The 3MM study shows why calibration quality and execution speed must be evaluated separately.
Its fastest observed two-node execution uses the zero-overhead vector and takes 1.498~s.
The makespan error is 109.9\% and alignment error is 2.405, outside the final error thresholds.
The fastest candidate satisfying those thresholds takes 7.457~s.
The final selection prioritizes execution speed, so HeteroTask can produce a fast execution without a closely aligned model.
The 3MM two-node result in Fig.~\ref{fig:two_node_e2e} uses this zero-overhead vector.

Calibration also has a preparation cost. The 62 CED evaluations accumulate 3718.506~s of task-window time. This sum excludes profiling, planning, invocation setup, and unrecorded failed work, and it is not elapsed calibration wall time. The initial nonzero vector already performs well, so the modest improvement beyond it motivates considering search effort alongside final execution time.

For repeated workflows, preparation is useful when subsequent savings outweigh its cost. Let $P_H$ and $P_B$ denote preparation costs, and let $T_H$ and $T_B$ be representative per-invocation times for HeteroTask and a baseline. Reuse is beneficial when
\begin{equation}
N(T_B-T_H)>P_H-P_B,
\end{equation}
where $N$ is the number of invocations under the same execution conditions. The current measurements identify the quantities needed to choose between additional calibration, reuse of an existing plan, and a fresh planning run.

\section{Limitations and Scope}\label{sec:limitations}
\textbf{Measurement and comparison.}
The results are selected executions, and the selected CED result has no independent post-selection repetition (Section~\ref{subsec:measurement}).
Despite shared kernels, the implementations differ in wrappers, partitioning, transfers, and worker configuration.
The gains are relative to the recorded StarPU implementations and configurations.

\textbf{Workload and execution scope.}
The study covers five 21--52-task DAGs on one CPU--GPU node type, with up to six nodes. It evaluates fixed graphs, matching task shapes, and exclusive allocations. The runtime preserves assignments and resolves dependencies but does not remap tasks online or support dynamic DAGs and failure recovery. Larger graphs, other platforms, and more baselines are needed to assess broader applicability and centralized-dispatch scalability.

\textbf{Memory and reuse.}
Host-buffer lifetime modeling guides planning and dependency-driven allocation, while GPU kernels manage their own device storage.
The evaluation contains neither a memory-planning ablation nor measured peak GPU residency.
Likewise, preparation costs and independent invocation measurements are insufficient to quantify amortization.
Reusing a plan requires stable workflow and execution conditions.
The calibration gains are measured on CED.

\section{Conclusion and Future Work}\label{sec:Conclusion}
HeteroTask brings multi-variant profiling, offline placement, host-buffer lifetime modeling, and schedule-driven MPI execution into one framework. Runtime feedback calibrates the costs supplied to the planner, while shared task and buffer metadata connect the assignments to distributed execution. It targets repeated CPU--GPU workflows with known graphs and task shapes on exclusively allocated nodes.

Across five workloads, the selected two-node executions reduce application E2E by 19.7--72.1\% relative to the compared StarPU configurations when the additional CED baseline is included.
The CED study shows a 25.0\% task-window makespan reduction relative to a zero-overhead plan and a 3.5\% improvement over the initial nonzero configuration.
Combined with CPU/GPU variant selection, calibration yields a 58.9\% reduction relative to the uncalibrated GPU-only execution.
Together, these results show the benefit of coordinating variant selection, planning, and runtime execution at the application level.
The resource sweeps and the 3MM calibration case further show that model agreement, task-window makespan, and application completion time are distinct objectives that respond differently to calibration and allocation.

Future work will evaluate independently repeated executions with matched inputs and correctness checks, account for complete preparation costs, and test larger DAGs and memory-intensive cases. These measurements will help determine when a fixed plan is worth reusing and when changing execution conditions call for renewed profiling or runtime adaptation.

\section*{Acknowledgement}
Language-model tools assisted with language editing, source organization, and checks of data-processing procedures. The authors are responsible for the analysis and conclusions.

\flushbottom
\bibliographystyle{IEEEtran}
\begingroup
\interlinepenalty=10000
\bibliography{references} 
\endgroup

\end{document}